\documentstyle[sprocl,12pt]{article}

\input epsf.tex

\def\ie{{\it i.e.\/}}

\def\anp#1#2#3{{\it Ann.\ Phys. (NY)} {\bf #1} (19#2) #3}
\def\arnps#1#2#3{{\it Ann.\ Rev.\ Nucl.\ Part.\ Sci.} {\bf #1}, (19#2) #3}
\def\cmp#1#2#3{{\it Comm.\ Math.\ Phys.} {\bf #1} (19#2) #3}
\def\ijmp#1#2#3{{\it Int.\ J.\ Mod.\ Phys.} {\bf A#1} (19#2) #3}
\def\jetp#1#2#3{{\it JETP Lett.} {\bf #1} (19#2) #3}
\def\jetpl#1#2#3#4#5#6{{\it Pis'ma Zh.\ Eksp.\ Teor.\ Fiz.} {\bf #1} (19#2) #3
[{\it JETP Lett.} {\bf #4} (19#5) #6]}
\def\jpb#1#2#3{{\it J.\ Phys.} {\bf B#1} (19#2) #3}
\def\mpla#1#2#3{{\it Mod.\ Phys.\ Lett.} {\bf A#1}, (19#2) #3}
\def\nci#1#2#3{{\it Nuovo Cimento} {\bf #1} (19#2) #3}
\def\npb#1#2#3{{\it Nucl.\ Phys.} {\bf B#1} (19#2) #3}
\def\plb#1#2#3{{\it Phys.\ Lett.} {\bf #1B} (19#2) #3}
\def\pla#1#2#3{{\it Phys.\ Lett.} {\bf #1A} (19#2) #3}
\def\prb#1#2#3{{\it Phys.\ Rev.} {\bf B#1} (19#2) #3}
\def\prc#1#2#3{{\it Phys.\ Rev.} {\bf C#1} (19#2) #3}
\def\prd#1#2#3{{\it Phys.\ Rev.} {\bf D#1} (19#2) #3}
\def\pr#1#2#3{{\it Phys.\ Rev.} {\bf #1} (19#2) #3}
\def\prep#1#2#3{{\it Phys.\ Rep.} {\bf C#1} (19#2) #3}
\def\prl#1#2#3{{\it Phys.\ Rev.\ Lett.} {\bf #1} (19#2) #3}
\def\rmp#1#2#3{{\it Rev.\ Mod.\ Phys.} {\bf #1} (19#2) #3}
\def\sjnp#1#2#3#4#5#6{{\it Yad.\ Fiz.} {\bf #1} (19#2) #3
[{\it Sov.\ J.\ Nucl.\ Phys.} {\bf #4} (19#5) #6]}
\def\zpc#1#2#3{{\it Zeit.\ Phys.} {\bf C#1} (19#2) #3}

\global\nulldelimiterspace = 0pt

\def\frac#1#2{{{#1} \over {#2}}\,}  
\def\hf{{1\over 2}}
\def\nth#1{{1\over #1}}

\def\del{\nabla}

\def\Dslsh{\hbox{/\kern-.6700em\it D}} 
\def\Scaslsh{\hbox{/\kern-.8200em${\cal A}$}} 
\def\dslsh{\hbox{/\kern-.5300em$\partial$}}
\def\pslsh{\hbox{/\kern-.5600em$p$}}
\def\sslsh{\hbox{/\kern-.5300em$s$}}
\def\epsslsh{\hbox{/\kern-.5100em$\epsilon$}}
\def\delslsh{\hbox{/\kern-.6300em$\nabla$}}
\def\lslsh{\hbox{/\kern-.4300em$l$}}
\def\elslsh{\hbox{/\kern-.4500em$\ell$}}
\def\kslsh{\hbox{/\kern-.5100em$k$}}
\def\qslsh{\hbox{/\kern-.5000em$q$}}
\def\slsh#1{\raise.15ex\hbox{$/$}\kern-.57em #1}

\def\roughly#1{\mathrel{\raise.3ex\hbox{$#1$\kern-.75em
   \lower1ex\hbox{$\sim$}}}}

\def\sss{\scriptscriptstyle}

\def\bfx{{\bf x}}

\def\Sca{{\cal A}}

\def\Sch{{\cal H}}
\def\Sci{{\cal I}}

\def\Scl{{\cal L}}

\def\Scr{{\cal R}}

\def\ssr{{\sss R}}

\def\Re{{\rm Re\;}}
\def\Im{{\rm Im\;}}

\def\eps{\epsilon}

\def\eq{\begin{equation}}
\def\eeq{\end{equation}}
\def\bg{\begin{eqnarray}}
\def\nd{\end{eqnarray}}
\def\nn{\nonumber}

\def\pref#1{(\ref{#1})}

\def\ie{{\it i.e.\/}}

\def\hf{{\frac12}}

\def\Re{{ \mbox{Re} \,}}
\def\Im{{ \mbox{Im} \,}}

\def\roughlyup#1{\mathrel{\raise.3ex\hbox{$\sim$\kern-.75em
\lower1ex\hbox{$#1$}}}}
\def\roughlydown#1{\mathrel{\raise.3ex\hbox{$#1$\kern-.75em
\lower1ex\hbox{$\sim$}}}}

\def\eqa{\begin{eqnarray}}
\def\eeqa{\end{eqnarray}}
\def\eq{\begin{equation}}
\def\eeq{\end{equation}}
\def\nn{\nonumber}

\def\eps{\epsilon}

\def\Sca{{\cal A}}

\def\sss{\scriptscriptstyle}

\begin{document}
\rightline{hep-ph/9812468}
\rightline{McGill-98/38}
\vspace{10mm}
\title{A Goldstone Boson Primer}
\author{C.P. Burgess}
\address{Physics Department, McGill University, 3600 University Street,\\
Montr\'eal, Qu\'ebec, Canada, H3A 2T8.}

\maketitle

\abstracts{These lectures are an extremely condensed 
version of the theory of Goldstone bosons, with general
features illustrated using a simple model. A more comprehensive 
version of these lectures, which includes a general discussion
of effective theories of Goldstone bosons,  
including applications to the low-energy behaviour of
pions, spin waves (in antiferromagnets and ferromagnets), and to
the $SO(5)$ proposal for high-$T_c$ superconductors may be
found in {\tt hep-th/9808176}.}

\section{Introduction}

George Bernard Shaw once observed that England and America
were divided by a common language. The same might be
said about the fields of theoretical high-energy, nuclear and 
condensed-matter physics. 
Since their joint start with the birth of quantum mechanics,
these three disciplines have diverged so far from one
another that it is very difficult for the practitioners of one
of these fields to follow in detail the developments and
techniques which are common in the others. This divergence
is unfortunate, since the cross-fertilization of ideas between
these fields has been a rich source of progress to all three.

And yet, their languages are very much the same.
There is, after all, considerable overlap in the 
theoretical techniques used in all three of these disciplines. 
On the broadest level (for very good reasons 
\cite{Weinberg96,Weinberg95/96}), 
all three heavily rely on field theory --- both classical and quantum. 
Other similarities also arise when they are inspected in more
detail, two of which play a significant role in these lectures. 

\begin{enumerate}
\item
All of these disciplines rely heavily on the appearance and utility 
of symmetries to analyze the behaviour of complicated processes. 
\item
All of these fields exploit low-energy expansions to take
advantage of the simplifications which often accompany large
heirarchies of scale. They also frequently use renormalization-group
techniques to resum singularities and identify scaling behaviour
away from characteristic energy scales. 
\end{enumerate}

The lectures summarized here describe a powerful theoretical 
technique which is based on the exploitation of symmetries 
and the simplicity of the low-energy limit, and so which has 
wide applications within the above-mentioned disciplines, as 
well as more widely throughout physics. The technique is the
use of effective field theories to describe low-energy behaviour,
specifically applied to the low-energy states --- Goldstone 
bosons --- which arise whenever a system's ground state
does not share all of the symmetries of its Hamiltonian.

For brevity's sake, the main ideas are presented here purely
within the context of a very simple model. The reader is
referred to ref.~\cite{Burgess98a} for all of the technical
details, including the general formulation of the low-energy
theory of Goldstone bosons as well as its application to
several examples from nuclear and condensed-matter physics. 
Although the model used is Lorentz invariant for simplicity,
the consequences drawn are not limited to this case.

\section{A Model}

Consider the model defined by the following Lagrangian density
for a complex scalar field, $\phi$:
\bg
\label{abeltoymodel}
\Scl &=& - \partial_\mu \phi^* \partial^\mu \phi - V(\phi^* \phi), \nn\\
\hbox{with} \qquad V &=& {\lambda \over 4} \; \left( \phi^* \phi - {\mu^2
\over \lambda} \right)^2 .
\nd

\subsection{Symmetries and Conservation Laws}

This theory describes two spinless particles which are related to
one another by a continuous $U(1)$ symmetry of the form 
\bg
\label{U1form}
\phi \to e^{i\omega} \; \phi. 
\nd 
The variation of $\Scl$ 
under an infinitesimal symmetry transformation, $\delta \phi
= i \omega \; \phi$,  is:
\bg
\label{deltaL}
\delta \Scl = - i \partial^\mu \phi \, \partial_\mu \omega, 
\nd
which shows that this transformation is a global
(or rigid) symmetry because it leaves the Lagrangian density 
invariant only if its parameter, $\omega$, is a constant. 

Continuous symmetries imply conservation laws. In ordinary
quantum mechanics the $U(1)$ symmetry considered here
would guarantee the time-independence (\ie\ conservation) 
of a hermitian charge operator $Q$ whose commutator with any operator
gives its transformation under the symmetry. In the present example:
\bg
\label{generator}
i \omega \Bigl[ Q, \phi(x) \Bigr] = \delta \phi(x) = i \omega \, \phi(x).
\nd

In field theories continuous symmetries carry an additional implication.
Besides implying the overall conservation of the charge $Q$, the
conservation laws must in addition hold locally. This implies 
the existence of a local (Noether) current, $j^\mu(x)$, 
for which conservation is the differential condition $\partial_\mu 
j^\mu = 0$. This expresses conservation because it implies the 
time independence of $Q$, which may be defined in terms of 
$j^\mu$ by $Q = \int j^0(\bfx,t) \; d^3\bfx$.
The Noether current for the model under consideration is:
\eq
\label{toync}
j_\mu = -i \left( \phi^* \partial_\mu \phi - \phi \partial_\mu \phi^* \right).
\eeq

\subsection{Semiclassical Ground State}

The next question is to determine the ground state and energy spectrum
in this model. This may be done explicitly if $\lambda \ll 1$ since
this condition justifies a semiclassical calculation of these quantities.

The field configuration which corresponds to the semiclassical 
ground state, or vacuum, is found by minimizing the system's 
energy density, which is $\Sch = \hf \, \partial_t{\phi}^* 
\partial_t{\phi} + \hf \, \del\phi^* \cdot \del\phi + V(\phi^*\phi)$. 
Being the sum of non-negative terms, this is easy to minimize. 
The vacuum configuration is found in this way to be a constant, 
$\dot\phi = \del\phi = 0$, whose value, $\phi = v$, minimizes the classical
potential: $V(v^*v) = 0$. Using the $U(1)$ symmetry to make 
$v$ real (and assuming $\mu^2$ is positive) gives the solution
$v = \mu /\sqrt{\lambda}$. 

The low-energy degrees of freedom in the semiclassical approximation 
consists of small harmonic oscillations of the fields about the minimum
of the scalar potential. The low energy spectrum is simply the
energy eigenvalues for each of these harmonic normal modes. 
Using the smallness of $\lambda$ to drop cubic and higher powers
of the fluctuation, $\phi - v$, and writing separately its real and 
imaginary part ($\Scr \equiv \sqrt{2} \; \Re (\phi - v)$ and 
$\Sci \equiv \sqrt{2} \; \Im \phi$) gives the harmonic 
Lagrangian $\Scl_{\rm h} =
-\hf \Bigl( \partial_\mu \Scr \partial^\mu \Scr + m_\ssr^2 \Scr^2 \Bigr)
-\hf  \; \partial_\mu \Sci \partial^\mu \Sci $, where $m_\ssr^2 
= \lambda v^2$. 

This gives the usual result: two particle types with a relativistic 
dispersion relation $E(p) = \sqrt{p^2 + m^2}$, with the particle
associated with the field $\Scr$ having rest mass $m_\ssr$ and
the particle associated with $\Sci$ have zero mass. 

\subsection{Particle Interactions}

For small $\lambda$ the interactions amongst these particles 
may be found perturbatively by expanding the scalar potential
in powers of $\Scr$ and $\Sci$:
\eq
\label{newabelpotl}
V = {m_\ssr^2 \over 2} \, \Scr^2 + {g_{30} \over 3!} \, \Scr^3 + {g_{12} \over 2} \, \Scr \Sci^2 + {g_{40} \over 4!} \, \Scr^4 + 
{g_{22} \over 4} \, \Scr^2 \Sci^2 + {g_{04} \over 4!} \, \Sci^4 ,
\eeq
where the couplings in this potential are given in terms of the
original parameters, $\lambda$ and $v$, by: 
\eq
\label{potcouplings}
{g_{30} \over 3!} = {g_{12} \over 2} =
{\lambda v \over 2 \sqrt{2}}, \qquad {g_{40} \over 4!} 
= {g_{04} \over 4!} = {g_{22} \over 8} = {\lambda \over 16}.
\eeq

An interesting point can be made if the amplitude for
$\Scr$--$\Sci$ scattering is computed to lowest-order in
perturbation theory using these interactions. The four
Feynman diagrams which contribute to this order are given
in Figure 1. 

\vspace{0.5cm}

\centerline{\epsfxsize=9.5cm\epsfbox{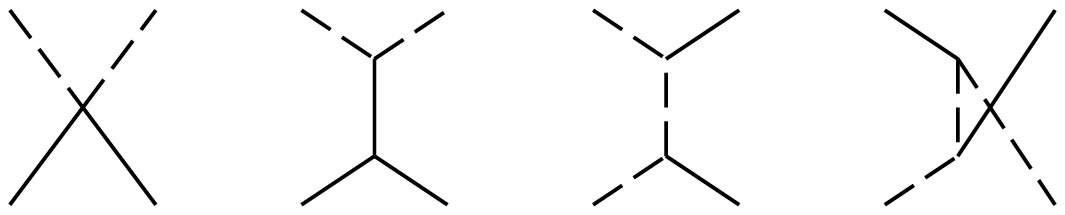}}

\begin{quote}
{\footnotesize               
{\bf Figure 1:} {\sl The Feynman graphs which describe $\Scr - \Sci$
scattering at tree level in this model.}}
\end{quote}


The $S$-matrix at tree which results from the evaluation of these
graphs is:
\eq
\label{smatrixdef}
S[\Scr(r) + \Sci(s) \to \Scr(r') + \Sci(s')] = {i \Sca \; 
\delta^4(r + s - r' - s') \over 4 (2\pi)^2 \sqrt{s^0 r^0 s'^0 r'^0}} \; ,
\eeq
with 
\eq
\label{smatrixresult}
\Sca = - g_{22} + { g_{12} g_{30} \over (s+s')^2 + m_\ssr^2 - i\eps} +
g_{12}^2 \left[ {1 \over (s+r)^2 -i \eps} + {1 \over (s - r')^2 - i\eps}
\right].
\eeq

An interesting feature of this amplitude is that it vanishes in the limit
of vanishing momentum for the massless particle, $\Sci$. That is,
(using the massive-particle dispersion relation, $r^2 = r'^2 = 
- m_\ssr^2$): 
\bg
\label{zeromomlim}
\Sca & \to & - g_{22} + { g_{12} g_{30} \over m_\ssr^2} - \,
{2 g_{12}^2 \over m_\ssr^2} , \nn\\
&=& \lambda \; \left( - \, \hf + {3 \over 2} \;  - 1 \right) = 0.
\nd

Even more interesting: as may be verified by more complicated 
calculations, the vanishing of $\Sca$ in the zero-momentum limit 
holds also for higher orders in perturbation theory. The same is
also true for all other amplitudes involving $\Sci$ particles. The
massless particle of this theory (which remains massless even once
interactions are included) completely decouples in the limit of 
vanishing momentum.

\section{The Low-Energy Perspective}

The purpose of this section is to show how the remarkable properties 
of the $\Sci$ particles just described can be exhibited more 
transparently, without resorting to detailed calculations.

The key idea is tha both the masslessness of $\Sci$ 
and its decoupling are properties
of the low-energy part of the model. They should be possible to
understand purely within the low-energy effective theory 
which is obtained by `integrating out' all degrees of freedom 
having energies higher than $O(m_\ssr)$ \cite{ETbooks,ETreviews}. 

\subsection{The Low-Energy Effective Lagrangian}

The interactions amongst $\Sci$ particles for centre-of-mass
energies much smaller than $m_\ssr$ can be described by
an `effective' Lagrangian, $\Scl_{\rm eff}(\Sci)$ 
which involves only the field $\Sci$. The field $\Scr$ does not 
appear in $\Scl_{\rm eff}$ because if no $\Scr$-particles
are initially present in a process, then they are also never 
produced (so long as $E_{\sss CM} < m_\ssr$) because
of energy conservation.  

That is not to say that $\Scr$ is irrelevant to low-energy $\Sci$-particle
scattering, however, because we know from the full theory
that virtual $\Scr$-exchange can and does take place. At low
energies its influence is suppressed by powers of $1/m_\ssr$,
because of the large energy denominators (or propagators)
which virtual $\Scr$ exchange requires. At low energies it is 
therefore useful to organize the terms in $\Scl_{\rm eff}$ in
powers of derivatives of $\Sci$ divided by $m_\ssr$, because
this completely captures the influence at low-energies of the
higher-energy components of the system.

The result of such a derivative expansion would be:
\bg
\label{dexp1}
\Scl_{\rm eff} = - V_{\rm eff}(\Sci) - \; \hf \, G(\Sci)
\, \partial_\mu \Sci \, \partial^\mu \Sci - \frac{H(\Sci)}{m^4}
\, (\partial_\mu \Sci \, \partial^\mu \Sci )^2 + \cdots,
\nd
where the ellipses describe further terms in the derivative
expansion. (Notice that the resulting lagrangian involves 
couplings with dimensions of inverse powers of mass --
in units for which $\hbar = c = 1$ -- and so is not
perturbatively renormalizable in the ordinary sense. It
nonetheless gives sensible predictions provided one works
to a fixed order in powers of $1/m_\ssr$.) 

The unknown functions $V_{\rm eff}$, $G$ and $H$ are 
determined by comparing $\Sci$-particle scattering computed
with $\Scl_{\rm eff}$ to the result computed within the full
model, with the Lagrangian of eq.~\pref{abeltoymodel}. 

\subsection{A Different Choice of Fields}

In the effective-Lagrangian language the masslessness of
$\Sci$ and the vanishing of all $S$-matrix elements in the 
zero-energy limit is equivalent to the vanishing of the effective 
potential: 
\bg
\label{Vis0}
V_{\rm eff} \equiv 0. 
\nd
The puzzle is to see in a simple way why this should be so.

To this end imagine instead using polar coordinates in field space,
rather than the fields $\Sci$ and $\Scr$:
\eq
\label{polcoords}
\phi(x) = \chi(x) \; e^{i \theta(x)} .
\eeq
In terms of $\theta$ and $\chi$ the model's Lagrangian is: 
\eq
\label{linpolcoords}
\Scl = - \partial_\mu \chi \partial^\mu \chi - \chi^2 \partial_\mu \theta
\partial^\mu \theta - V(\chi^2).
\eeq
Analyzing the spectrum of this theory in the semiclassical approximation
about the vacuum $\chi = v$ shows that $\chi$ describes the particle
with mass $m_\ssr$ and $\theta$ represents the massless particle. 

Now comes the main point. The field $\theta$ only appears in $\Scl$
through its derivative, $\partial_\mu \theta$. Suppose we were now
to integrate out the degrees of freedom having energies of order
$m_\ssr$. In the resulting effective lagrangian $\theta$ must also
appear only differentiated. Using these variables we therefore 
easily see that $V_{\rm eff}(\theta) \equiv 0$, and so why the
massless particle decouples at low energy.

\subsection{A Tradeoff}

Weinberg's First Law of Theoretical Physics states 
\cite{WeinbergGL}: You can use any variables at all to 
analyze a problem, but if you use the wrong variables you'll
be sorry.

In this problem the massless particles decouple at low energy
regardless of whether $\Scl$ is written using the fields $\Sci$
and $\Scr$ or the fields $\chi$ and $\theta$. The latter pair
have the advantage that they display the low-energy decoupling
of $\theta$-particles in a transparent way, and so they more
clearly exhibit the limits of validity of this decoupling.

There is a price for this clarity, however. This price is most 
easily seen once the fields are canonically normalized, 
which is acheived by writing $\chi = v +
\nth{\sqrt{2}} \; \chi'$ and $\theta = \nth{v \sqrt{2}} \; \varphi$. With
these variables the Lagrangian is seen to have acquired complicated,
nominally nonrenormalizable interactions:
\eq
\label{nrints}
\Scl_{\rm nr} = - \left[ {\chi' \over \sqrt{2} \; v^2} + {\chi'^2 \over 4 v^2}
\right] \; \partial_\mu \varphi \partial^\mu \varphi. 
\eeq
Notice this lagrangian only makes sense with this choice of variables
if $v \ne 0$.

Of course, the $S$-matrix for the theory in these variables is identical to
that derived from the manifestly renormalizable Lagrangian expressed in terms
of the variables $\Scr$ and $\Sci$. So the $S$-matrix remains renormalizable
even when computed using the variables $\chi'$ and $\varphi$. The same is not
true for {\em off-shell} quantities like Green's functions or the 1PI action,
however, since the renormalizability of these quantities need not survive a
nonlinear field redefinition. 

In this model there is a choice to be made between making the
Lagrangian manifestly display either the renormalizability of the theory, or
the Goldstone boson nature of the massless particle. Which is best to keep
explicit will depend on which is more convenient for the calculation that is
of interest. Since, renormalizability is in any case given up
when dealing with effective low-energy field theories, it is clear that the
variables which keep the Goldstone boson properties explicit are the ones of
choice in this case. 

\section{Naturalness and Goldstone Bosons}

Although use of the variables $\chi$ and $\theta$ clearly
display the special properties of the massless particle, it is not
yet clear {\em why} these special properties arise in the first
place. This section addresses this issue, first by identifying
the $U(1)$ symmetry within the low-energy effective theory.
The resulting symmetry argument is then shown to be a the
low-energy expression of an exact result: Goldstone's theorem.

\subsection{Symmetry Considerations}

The key to understanding the properties of the massless
particle lie with the model's $U(1)$ symmetry, eq.~\pref{U1form}.
This is realized on the fields $\Sci$ and $\Scr$ as a two-by-two 
orthogonal rotation, but it is realized on the fields $\chi$ and
$\theta$ {\em inhomogeneously}. In terms of the canonically normalized
field, $\varphi$, this transformation law becomes: 
\eq
\label{gbabeltransf}
\varphi \to \varphi + \sqrt{2} \; v \; \omega. \eeq

Clearly it is this symmetry which requires $\theta$ to
appear only differentiated in both $\Scl$ and $\Scl_{\rm eff}$,
because only $\partial_\mu \phi$ is invariant with respect to
constant shifts of $\phi$. This symmetry therefore is also at
the root of the masslessness and low-energy decoupling of
the particle described by $\theta$. 

\subsection{Spontaneously-Broken Symmetries}

An inhomogeneous symmetry of the form $\theta \to \theta
+ \omega$ is indicative of a symmetry which does not 
preserve the system's ground state. Such a symmetry
is called {\sl spontaneously broken}. 

Recall that if $\chi$ is frozen to equal its value in the
vacuum, $\chi = v \ne 0$, then a nonzero field configuration 
for $\theta$ corresponds to $\phi(x) = v \; \exp[ i \theta(x)]$. 
In this sense $\theta$ can be regarded as being the result of
performing a spacetime-dependent $U(1)$ transformation 
of the ground state. But this $U(1)$ is a symmetry only when
its parameter is a constant, and $\Scl$ vanishes when
evaluated at the vacuum configuration, $\phi = v$. We see
that $\Scl$ is independent of $\theta$ as $\theta$ tends to
a constant, precisely because $\theta$ is simply a symmetry
transformation of the vacuum in this limit. 

If $\theta$ varies in space and time it no longer describes a
symmetry transformation, and so $\Scl$ can depend on 
derivatives of $\theta$. So it is continuity with the constant-field
limit, where $\theta$ parameterizes a symmetry transformation,
which ensures the low-energy decoupling of $\theta$.

\subsection{Goldstone Bosons}

The argument just given is very general. It states that for
{\sl any} continuous symmetry which does not preserve the
ground state, there is a massless degree of freedom which
decouples at low energies. This mode is called the Goldstone
(or Nambu-Goldstone) particle for the symmetry. 

Furthermore, this degree of
freedom may be explicitly displayed by performing the
symmetry transformation in question on the ground-state
field configurations. If the parameters of this transformation
are treated as fields they automatically drop out of the
Lagrangian (or effective Lagrangian) in the limit of constant
fields. 

This observation implies a number of other consequences
for the Goldstone particles, in addition to their masslessness
and low-energy decoupling \cite{Guralnik68,LeutwylerGB,Hofmann}:

\begin{enumerate}
\item
{\it Spin and Statistics:}
For internal symmetries the Goldstone particles are spinless bosons,
since they can be represented by fields which are rotational scalars
(\ie\ the transformation parameters of the symmetry group). An
identical argument for spontaneously-broken supersymmetry 
implies the Goldstone particles are spin-half fermions, and they
are spin-one bosons (phonons)
for spontaneously-broken translation invariance.
\item
{\it Counting:}
There is precisely one Goldstone particle for each symmetry
generator which is broken. For example, if $U(N)$ (which
has $N^2$ generators) is spontaneously broken to $U(N')$
then there must be $N^2 - {N'}^2$ Goldstone bosons. 
\item
Finally, applying the symmetry transformation of eq.~\pref{gbabeltransf}
to the Goldstone boson kinetic terms, $-\, \hf \, \partial_\mu \varphi
\, \partial^\mu \varphi$, implies the corresponding conserved
current depends on the Goldstone boson in the following way: 
\eq
\label{abelgbncex}
j^\mu = \sqrt{2} \; v \; \partial^\mu \varphi + \cdots \, . 
\eeq
The ellipses in this expression represent contributions to $j^\mu$ which come
from other terms in the Lagrangian besides the $\varphi$ kinetic term,
and so involve other fields or additional derivatives of powers of
$\varphi$. 

Eq.~\pref{abelgbncex} implies another general property of Goldstone
bosons: the matrix element of the current between the ground state,
$|\Omega\rangle$ and the Goldstone state, $|G\rangle$, must
be nonzero: $\langle{G}| j^\mu | \Omega\rangle \neq 0$. 
\end{enumerate}

\section{Discussion}

The model examined here illustrates a general property of
field theories. When a continuous, global symmetry is
spontaneously broken, the spectrum must contain a massless
(Goldstone) particle which completely decouples at zero energy, and so
is weakly-interacting at low energies. 

The special low-energy properties of Goldstone bosons are all
consequences of their particular form of inhomogeneous transformation
law under the corresponding broken symmetry. (For abelian 
symmetries this transformation rule is as in eq.~\pref{gbabeltransf},
but for nonabelian symmetries a more complicated form
is required \cite{ChiPT}.) Because 
Goldstone-boson properties all can be derived 
purely on the grounds of their symmetry transformation properties, 
they do not depend at all (at low energies) on the details of the 
underlying model which breaks these symmetries. 

More generally, dependence on the underlying model appears
once predictions are required beyond the leading order in the
low-energy derivative expansion. But even once subleading
corrections are included, underlying physics affects Goldstone
boson interactions only through a comparatively small number 
of parameters. 

To show that this is true, imagine writing an arbitrary effective theory for a
real scalar field, $\varphi$, subject only to the symmetry of
eq.~\pref{gbabeltransf} (and, for simplicity, to Poincar\'e invariance).  The
most general Lagrangian which is invariant under this transformation is an
arbitrary function of the derivatives, $\partial_\mu\varphi$, of the field.
An expansion in interactions of successively higher dimension 
would be:
\eq
\label{abelgbaction}
\Scl_{\rm eff}(\varphi) = - \hf \; \partial_\mu \varphi 
\partial^\mu \varphi - {a \over
v^4} \;  \partial_\mu \varphi \partial^\mu \varphi \; \partial_\nu \varphi
\partial^\nu \varphi + \cdots,
\eeq
where a power of $v$ is inserted on as appropriate to ensure that the
parameter $a$ is dimensionless. This accords with the expectation
that it is the symmetry-breaking scale, $v$, which sets the natural scale
relative to which the low energy limit is to be taken. In the example
considered earlier, integrating out the heavy field, $\chi'$ 
produces these powers of $v$ through the appearance of 
inverse  powers of $m_\ssr$. 

It follows that, up to subleading order in $1/m_\ssr$, the 
mutual scattering of Goldstone bosons in {\sl any}
model which spontaneously breaks $U(1)$ depends on the
details of the model only through its predictions for the
parameter $a$. This is because the integrating out of all
other, heavier, degrees of freedom necessarily must give
an effective Lagrangian of the form of eq.~\pref{abelgbaction}, but
with a specific, calculable coefficient for the parameter $a$.

Such an understanding of the Goldstone nature of a field, like $\varphi$, as
an automatic consequence of a symmetry is clearly invaluable when constructing
effective Lagrangians for systems subject to spontaneous symmetry breaking.

\section*{Acknowledgements}
I am indebted to Prof. Dong-Pil Min and the organizers of
the Nuclear Physics Summer School and Symposium (NuSS'98) 
for the kind invitation to give these lectures, as well as the warm
hospitality extended to me during me stay at Seoul National University.

These lectures are partially based on a longer series of lectures
given in E.T.H. Lausanne and the Universit\'e de
Neuch\^atel in June 1995, entitled 
{\it An Introduction to Effective Lagrangians and Their Applications}. 

My principal research funds come from {\sl the Natural Sciences and 
Engineering Council of Canada}, with some additional funds 
being provided by {\sl les Fonds pour la Formation de Chercheurs 
et l'Aide \`a la R\'echerche du Qu\'ebec}.


\def\anp#1#2#3{{\it Ann.\ Phys. (NY)} {\bf #1} (19#2) #3}
\def\apj#1#2#3{{\it Ap.\ J.} {\bf #1}, (19#2) #3}
\def\arnps#1#2#3{{\it Ann.\ Rev.\ Nucl.\ Part.\ Sci.} {\bf #1}, (19#2) #3}
\def\cmp#1#2#3{{\it Comm.\ Math.\ Phys.} {\bf #1} (19#2) #3}
\def\ejp#1#2#3{{\it Eur.\ J.\ Phys.} {\bf #1} (19#2) #3}
\def\ijmp#1#2#3{{\it Int.\ J.\ Mod.\ Phys.} {\bf A#1} (19#2) #3}
\def\jetp#1#2#3{{\it JETP Lett.} {\bf #1} (19#2) #3}
\def\jetpl#1#2#3#4#5#6{{\it Pis'ma Zh.\ Eksp.\ Teor.\ Fiz.} {\bf #1} (19#2) #3
[{\it JETP Lett.} {\bf #4} (19#5) #6]}
\def\jpa#1#2#3{{\it J.\ Phys.} {\bf A#1} (19#2) #3}
\def\jpb#1#2#3{{\it J.\ Phys.} {\bf B#1} (19#2) #3}
\def\mpla#1#2#3{{\it Mod.\ Phys.\ Lett.} {\bf A#1}, (19#2) #3}
\def\nci#1#2#3{{\it Nuovo Cimento} {\bf #1} (19#2) #3}
\def\npb#1#2#3{{\it Nucl.\ Phys.} {\bf B#1} (19#2) #3}
\def\plb#1#2#3{{\it Phys.\ Lett.} {\bf #1B} (19#2) #3}
\def\pla#1#2#3{{\it Phys.\ Lett.} {\bf #1A} (19#2) #3}
\def\pra#1#2#3{{\it Phys.\ Rev.} {\bf A#1} (19#2) #3}
\def\prb#1#2#3{{\it Phys.\ Rev.} {\bf B#1} (19#2) #3}
\def\prc#1#2#3{{\it Phys.\ Rev.} {\bf C#1} (19#2) #3}
\def\prd#1#2#3{{\it Phys.\ Rev.} {\bf D#1} (19#2) #3}
\def\pr#1#2#3{{\it Phys.\ Rev.} {\bf #1} (19#2) #3}
\def\prep#1#2#3{{\it Phys.\ Rep.} {\bf #1} (19#2) #3}
\def\prl#1#2#3{{\it Phys.\ Rev.\ Lett.} {\bf #1} (19#2) #3}
\def\prs#1#2#3{{\it Proc.\ Roy.\ Soc.} {\bf #1} (19#2) #3}
\def\rmp#1#2#3{{\it Rev.\ Mod.\ Phys.} {\bf #1} (19#2) #3}
\def\sjnp#1#2#3#4#5#6{{\it Yad.\ Fiz.} {\bf #1} (19#2) #3
[{\it Sov.\ J.\ Nucl.\ Phys.} {\bf #4} (19#5) #6]}
\def\zpc#1#2#3{{\it Zeit.\ Phys.} {\bf C#1} (19#2) #3}


\end{document}